\documentclass[10pt,twoside,a4paper]{article} 
\usepackage{rgd21pap}
\begin{document}
\setcounter{page}{1}
%%%%%%%%%%%%%%%%%%%%%%%%%%%%%%%%%%%%%%%%%%%%%%%%%%%%%%%%%%%%%%%%%
% INDICATE BELOW THE DATA TO BE USED FOR CORRESPONDENCE RELATIVE
% TO THIS PAPER. REPLACE 0000 BY YOUR ACTUAL ABSTRACT NUMBER
%
%paper/abstract number: 5377
%name of person to contact: Andr\'es Santos
%
%%%%%%%%%%%%%%%%%%%%%%%%%%%%%%%%%%%%%%%%%%%%%%%%%%%%%%%%%%%%%%%%%%
% Insert the full title of your paper in the title command:
% Capitalize the first character of each significant word
\title{Non-Newtonian Shear Viscosity \\
in a Dense System of Hard Disks}

%%%%%%%%%%%%%%%%%%%%%%%%%%%%%%%%%%%%%%%%%%%%%%%%%%%%%%%%%%%%%%%%%
% Insert the authors' names and affiliations as shown below:
% Use exponents only if authors have different affiliations
% For each author, write initials first, then family name
% Do not use "and" to introduce the last author
% Use simplified addresses (typically: laboratory, organization, town, country)

\author{J.M. Montanero$^1$, A. Santos$^2$\\
$^1$ Departamento de Electr\'onica e Ingenier\'{\i}a Electromec\'anica,\\
Universidad de Extremadura, E-06071 Badajoz, Spain\\
$^2$ Departamento de F\'{\i}sica,\\ Universidad de Extremadura, E-06071 
Badajoz, Spain
}

%%%%%%%%%%%%%%%%%%%%%%%%%%%%%%%%%%%%%%%%%%%%%%%%%%%%%%%%%%%%%%%%%
% Do not use the \date command
\maketitle 

%%%%%%%%%%%%%%%%%%%%%%%%%%%%%%%%%%%%%%%%%%%%%%%%%%%%%%%%%%%%%%%%%
% Type in your PAPER, starting below:

\section{Introduction}
\label{p5377sec1}
Kinetic theory can be viewed as an intermediate step  between a
detailed microscopic analysis of a many-body system and 
the corresponding phenomenological macroscopic description. 
In  kinetic theory the 
main objective is to derive and solve the kinetic equation 
for the one-particle distribution function, thus obtaining information about 
the system
 properties.
In the context of dilute gases, the Boltzmann equation (BE) 
\cite{p5377r1} provides the
adequate framework for studying states arbitrarily far from equilibrium.
Exact solutions to this equation are rare, but a great deal of 
information  can be obtained from simplified kinetic models \cite{p5377r2}
or from simulation Monte Carlo  methods \cite{p5377r3}. 
On the other hand, the assumptions implicit in the BE are
only physically justified in the low-density limit. As the density increases,
structural effects  become important, potential  contributions to the fluxes
dominate, and the BE is no longer adequate. There is
no general kinetic equation valid for finite densities.
A singular exception, however, is the idealized system of hard spheres
of diameter $\sigma$, for which Enskog 
proposed a semi-phenomenological 
equation \cite{p5377r1} by introducing two crucial changes 
in the Boltzmann collision integral: (a) the centers of two colliding 
particles are separated by a distance equal to $\sigma$; 
(b) the collision frequency is increased by a factor that accounts for 
the spatial correlation between the two colliding molecules.
Although the Enskog equation (EE) also ignores the correlations in the 
velocities
before collision ({\em stosszahlansatz}), it leads to transport coefficients 
that are in  good
agreement with experimental and simulation values for a 
wide range of densities. In addition, 
the revised Enskog theory (RET) \cite{p5377r4}
%developed by van Beijeren and Ernst, 
is asymptotically exact at short
times and therefore has no limitations on density or space scale in that 
limit. Moreover, it admits both fluid and crystal equilibrium states as 
stationary
solutions.

The mathematical complexity of the EE has hindered practical applications.
As in the case of BE, two different approaches have been 
proposed to cope with this problem. First, a  Monte Carlo 
algorithm has been introduced to solve numerically the EE \cite{p5377r5}, 
in the same spirit as the DSMC method of solving the BE \cite{p5377r3};
second, a simple kinetic model that retains 
the main features of the EE has been constructed \cite{p5377r6,p5377r6bis}; 
it reduces in the low density limit
to the simplest kinetic model of the BE, the
Bhatnagar-Gross-Krook (BGK) model.
Both approaches have demonstrated to succeed in capturing the essential
properties of the EE and have great potential
for a new understanding of nonequilibrium systems under conditions 
accessible previously only by molecular dynamics simulation.

Two-dimensional systems often serve as prototypes to investigate some 
physical properties also present in real systems.
In particular, many of the peculiarities of a hard-sphere fluid in 
far from equilibrium states are expected to be present in a 
hard-disk system with a similar value of the packing fraction.  
Obviously, the calculations usually
become easier from both  theoretical and  computational points of view.
In this paper we calculate the rheological properties 
of a dense fluid of hard disks under shear far from equilibrium by
using a kinetic model of the EE.
The results are compared with Monte Carlo simulations. 
As happens in the three-dimensional case \cite{p5377r6bis}, the comparison
shows an excellent agreement at all densities and shear rates considered.

\section{The  Enskog Equation for a Hard-Disk Fluid under Uniform Shear Flow}
\label{p5377sec2}
The uniform shear flow (USF) is one of the few inhomogeneous states for which
exact results can be obtained far from equilibrium, and therefore is
of great significance in providing insight for the type of phenomena that
occur under extreme state conditions. Moreover, it has been studied
extensively by molecular dynamics simulation to analyze rheological 
properties in simple atomic fluids. 
The macroscopic state is characterized by a constant density $n$, 
a uniform temperature $T$, and a linear flow field:  
${\bf u}({\bf r})={\sf a}\cdot {\bf r}=ay\widehat{{\bf x}}$, 
where ${\sf a}=a \widehat{{\bf x}}\widehat{{\bf y}}$. The (constant) 
shear rate $a$ is a single parameter that can be chosen to drive the system 
arbitrarily far from equilibrium. The shear produces viscous heating that is 
compensated by an external nonconservative force (thermostat),
${\bf F}=-m\alpha(a){\bf V}$, where $m$ is the mass of a particle, ${\bf 
V}={\bf v}-{\bf u}$ is the peculiar velocity,
and the thermostat parameter $\alpha$
is adjusted to assure that the temperature remains constant. 
At a microscopic level, the USF is characterized by a distribution function,
$f({\bf r},{\bf v},t)\to f({\bf V},t)$, that becomes uniform in the 
Lagrangian frame of reference.
Under the above conditions, the EE for $f({\bf V},t)$ becomes
\begin{equation}
\label{p5377e1}
\left( \frac{\partial}{\partial t}-aV_y
\frac{\partial}{\partial V_x}-\alpha \frac{\partial}{\partial {\bf V}}
\cdot {\bf V} \right)f=J_E[f]\; ,
\end{equation}
where $J_E[f]$ is the Enskog collision operator,
\begin{equation}
\label{p5377e2}
J_E[f]= \sigma \chi(n)\int d{\bf V}_1\int d\widehat{
\mbox{\boldmath{$\sigma$}}}\
\Theta(\widehat{\mbox{\boldmath{$\sigma$}}}\cdot {\bf g})\ 
(\widehat{\mbox{\boldmath{$\sigma$}}}\cdot {\bf g})
[f({\bf V}',t)f({\bf V}_1^{'},t)-f({\bf V},t)f({\bf V}_1,t)]\; .
\end{equation}
In the above expression $\sigma$ is the disk diameter, $\chi(n)$
is the pair correlation function at contact for an equilibrium system with
(uniform) density $n$, $\widehat{\mbox{\boldmath{$\sigma$}}}$
is a unit vector, $\Theta(x)$ is the Heaviside function,
${\bf g}={\bf V}-{\bf V}_1-\sigma {\sf a}\cdot
\widehat{\mbox{\boldmath{$\sigma$}}}$,
${\bf V}^{'}={\bf V}-
(\widehat{\mbox{\boldmath{$\sigma$}}}\cdot {\bf g})
\widehat{\mbox{\boldmath{$\sigma$}}}$,
and ${\bf V}^{'}_1={\bf V}_1+2 \sigma {\sf a}\cdot
\widehat{\mbox{\boldmath{$\sigma$}}}+
(\widehat{\mbox{\boldmath{$\sigma$}}}\cdot
{\bf g})\widehat{\mbox{\boldmath{$\sigma$}}}$.

The most relevant transport quantity is the steady-state pressure tensor ${\sf 
P}$, which 
measures shear and normal stresses. It has a kinetic part, ${\sf P}^k$, and 
a collisional transfer part, ${\sf P}^c$, that are functionals of $f$ given 
by
\begin{equation}
\label{p5377e4.0}
{\sf P}^k=m\int d{\bf V} \,
{\bf V}{\bf V}
f({\bf V})\; ,
\end{equation}
\begin{equation}
\label{p5377e4}
{\sf P}^c=\frac{m\sigma^2 \chi}{2}\int d{\bf V} \int d{\bf V}_1
\int d\widehat{\mbox{\boldmath{$\sigma$}}} \
\widehat{\mbox{\boldmath{$\sigma$}}}
\widehat{\mbox{\boldmath{$\sigma$}}}\
\Theta(\widehat{\mbox{\boldmath{$\sigma$}}}\cdot {\bf g})
(\widehat{\mbox{\boldmath{$\sigma$}}}\cdot {\bf g})^2 f({\bf V}
+\sigma {\sf a}\cdot
\widehat{\mbox{\boldmath{$\sigma$}}})
f({\bf V}_1)\; .
\end{equation}

Following the standard Chapman-Enskog method \cite{p5377r1}, 
the Navier-Stokes constitutive equations are derived 
and the Newtonian shear viscosity can be identified as \cite{p5377r8}
\begin{equation}
\label{p5377e0}
\eta_{\mbox{\scriptsize{{NS}}}}(n)=\frac{\eta_0}{\chi}
\left(1+\frac{\pi}{4}n^*\chi\right)^2+\frac{1}{4\sigma}
{n^*}^2 \chi(\pi mk_BT)^{1/2}\; ,
\end{equation} 
where $\eta_0=1.022(mk_BT/\pi)^{1/2}/2\sigma$ is the Boltzmann viscosity,
$k_B$ is the Boltzmann constant,
and $n^*=n\sigma^2$ is the reduced density.
This Navier-Stokes shear viscosity is the zero shear rate limit ($a\to 0$)
of a generalized  transport coefficient
$\eta(n,a)=-P_{xy}/a$.
Other non-Newtonian effects are associated with the differences
$P_{xx}(n,a)-p_0(n)$ and $P_{yy}(n,a)-p_0(n)$, where $p_0(n)=
nk_BT(1+\frac{\pi}{2} n^*\chi)$ is the equilibrium hydrostatic pressure.

\section{The Kinetic Model}
\label{p5377sec3}
Very recently, a kinetic model 
has been derived by replacing the Enskog collision operator 
with a simpler form that, otherwise, retains  the main qualitative features.
The model has the same domain of applicability and preserves the same basic
properties
(such as local conservation laws 
and the exact equilibrium stationary state for
both fluid and crystal phases)
as the RET.
For a detailed account of the kinetic model, we refer the reader
to Refs.\ \cite{p5377r6,p5377r6bis}.
In the particular case of the USF,  the kinetic model leads
to the replacement 
\begin{equation}
\label{p5377e3}
J_E[f]\rightarrow -\nu(f-f_{\ell})-
f_{\ell}\left[\frac{P_{xy}^{c}}{nk_BT}a \left(\frac{m}{2k_BT}V^2-1\right)-
\frac{2m}{k_BT}V_xV_y A_{xy}\right]\; ,
\end{equation}
where we have already considered the two-dimensional case. In
Eq.\ (\ref{p5377e3})
$\nu$ represents an effective collision frequency depending on
the local density and temperature. Here, this parameter
is chosen to assure that the low density shear viscosity 
is the same as that from the BE, $\nu=nk_BT\chi/\eta_0$.
In addition, $f_{\ell}({\bf V})$ is the 
local equilibrium distribution,  $P_{xy}^c$ is the $xy$-element 
of the collisional transfer pressure tensor, as obtained from Eq.\
(\ref{p5377e4}),
and $A_{xy}$ is the collisional moment
\begin{equation}
\label{p5377e5}
A_{xy}=\frac{m}{2nk_BT} \int d{\bf V} V_x V_y J_E[f_{\ell}]
=-\frac{\pi}{4} (k_BT/m)^{\frac{1}{2}}n\sigma \chi \overline{a}
\left(1+\frac{3}{8} \overline{a}^2\right)\; ,
\end{equation}
where $\overline{a}\equiv \frac{1}{2}a\sigma(m/k_BT)^{1/2}$.
Equation (\ref{p5377e1})
together with the substitution (\ref{p5377e3}) constitutes now
the kinetic equation for the problem.
Since the term $P_{xy}^c$ is a functional of $f$,
Eq.\ (\ref{p5377e1}) [along with (\ref{p5377e3})]
is still a highly nonlinear integro-differential equation. 

In order to ease the notation, we choose units such that
$\nu=1$, $m=1$, and $2k_BT/m=1$, and define the dimensionless
pressure tensor ${\sf P}^*\equiv{\sf P}/nk_BT$. 
In these units, $\sigma=(\sqrt{2\pi}/1.022)n^*\chi$. 
Conservation of energy gives the thermostat parameter  
 $\alpha(a)$
in terms of $P_{xy}^*(a)$: 
\begin{equation}
\label{p5377e6}
\alpha(a)=-\frac{a}{2} P_{xy}^*(a)\; .
\end{equation}
Taking moments in both
sides of the (stationary) kinetic equation, one can easily get the kinetic 
part of the
pressure tensor:
\begin{equation}
\label{p5377e8}
P_{xx}^{k*}=
%2-P_{yy}^{k*}=
1+\frac{a^2-2 a A_{xy}}{
(1+2\alpha)^2+a^2}\; ,\quad
%\end{equation}
%\begin{equation}
%\label{p5377e7}
P_{xy}^{k*}=
\frac{1+2\alpha}{(1+2\alpha)^2+a^2}(2A_{xy}-a) \; .
\end{equation}
In order to close the mathematical problem, we would need to express
$P_{xy}^{c*}$ in terms of $\alpha$. In principle, this implies to perform
the velocity integrals (\ref{p5377e4})
with the formal solution to the kinetic
equation. Instead, we obtain a reasonable estimate for $P_{xy}^{c*}$ by
using the first Sonine approximation
\begin{equation}
\label{p5377e9}
f\rightarrow f_{\ell}\left[1+
(V_x^2-V_y^2)(P_{xx}^{k*}-1)+2
V_x V_y P_{xy}^{k*}\right]\; .
\end{equation}
With this approximation, the evaluation of ${\sf P}^{c*}$ is similar to 
that
of $A_{xy}$. In particular,
\begin{eqnarray}
\label{p5377e10}
P_{xy}^{c*}&=&-\frac{n^*\chi}{2}
\int d\widehat{\mbox{\boldmath{$\sigma$}}}\
\widehat{\sigma}_x \widehat{\sigma}_y
\left\{
(1+2 \overline{a}^2 \widehat{\sigma}_x^2
\widehat{\sigma}_y^2)\ \mbox{erf}(\overline{a} \widehat{\sigma}_x
\widehat{\sigma}_y) 
-2 \widehat{\sigma}_x 
\widehat{\sigma}_y P_{xy}^{k*} \right.
\nonumber \\
&&\left.+ \frac{\overline{a} \widehat{\sigma}_x
\widehat{\sigma}_y}{4\sqrt{\pi}}
e^{-\overline{a}^2 \widehat{\sigma}_x^2
\widehat{\sigma}_y^2} 
\left[8-\left(2\widehat{\sigma}_x^2-1\right)^2 
(P_{xx}^{k*}-1)^2-4 \widehat{\sigma}_x^2\widehat{\sigma}_y^2 
P_{xy}^{k*2}\right]\right\}
\; .\nonumber \\
\end{eqnarray}
From Eqs.\ (\ref{p5377e6}), (\ref{p5377e8}) and
(\ref{p5377e10}) one gets a {\em closed\/} equation for $\alpha$, 
whose numerical solution can be easily obtained for 
arbitrary shear rates and densities. 

\section{The ESMC Method}

As discussed in the Introduction, a recent method has been developed
for Monte Carlo simulation of the solution 
to the EE  \cite{p5377r5}.
Previous results have demonstrated the utility of this Enskog 
simulation Monte Carlo (ESMC) method for
studying far from equilibrium states in the regime of low and moderate
densities. In these conditions, the ESMC method  
can be even more efficient, from a computational point of view,
than hard-sphere molecular dynamics. In addition, the ESMC 
algorithm
reduces to the well-known DSMC method \cite{p5377r3} in the low density limit.

As applied to the USF, the method proceeds as follows. The one-particle
distribution function $f({\bf V})$ is represented by the peculiar 
velocities $\{ {\bf V}_i\}$ of a sample of $N$ ``simulated" particles. These
velocities are updated at integer times $t=\Delta t, 2\Delta t,
3\Delta t,\ldots$, where the time-step $\Delta t$ is much smaller than the
mean free time and the inverse shear rate. This is done in two stages:
free streaming and collisions. The free streaming stage consists  of
making 
\begin{equation}
\label{p5377e11}
{\bf V}_i\rightarrow e^{-\alpha \Delta t}({\bf V}_i-{\sf a}\cdot
{\bf V}_i \Delta t)\; ,
\end{equation}
where the thermostat parameter $\alpha$ is adjusted to assure that
the temperature, which is computed as $T=(m/Nk_B)\sum_i V_i^2$, remains 
constant. In the collision stage, a sample of $\frac{1}{2}N\widetilde{w}$ 
pairs are chosen at random with equiprobability, where $\widetilde{w}$ is an 
upper bound estimate of the
probability that a particle collides in the time interval between
$t$ and $t+\Delta t$. For each pair $ij$ belonging to this sample,
the following steps are taken: (1) a given direction 
$\widehat{\mbox{\boldmath{$\sigma$}}}_{ij}$
is chosen at random with equiprobability;
(2) the collision between particles $i$ and $j$ is accepted with a
probability equal to the ratio
$w_{ij}/\widetilde{w}$, 
where $w_{ij}=2\pi \sigma \chi n  \Delta t
\Theta(\widehat{\mbox{\boldmath{$\sigma$}}}_{ij}\cdot {\bf g}_{ij})
(\widehat{\mbox{\boldmath{$\sigma$}}}_{ij}\cdot {\bf g}_{ij})$ 
and ${\bf g}_{ij}={\bf V}_{i}-{\bf V}_j-\sigma {\sf a}\cdot
\widehat{\mbox{\boldmath{$\sigma$}}}_{ij}$;
and (3) if the collision is accepted, post-collision velocities are
assigned to both particles:
\begin{equation}
\label{p5377e12}
{\bf V}_i\rightarrow {\bf V}_i-
\widehat{\mbox{\boldmath{$\sigma$}}}_{ij}
(\widehat{\mbox{\boldmath{$\sigma$}}}_{ij}\cdot {\bf g}_{ij})\; ,\quad
%\end{equation}
%\begin{equation}
%\label{p5377e13}
{\bf V}_j\rightarrow {\bf V}_j+\widehat{\mbox{\boldmath{$\sigma$}}}_{ij}
(\widehat{\mbox{\boldmath{$\sigma$}}}_{ij}\cdot {\bf g}_{ij})\; .
\end{equation}
In the case that in one of the collisions 
$w_{ij}>\widetilde{w}$, the estimate of $\widetilde{w}$ is
updated as $\widetilde{w}=w_{ij}$. In the course of the simulations,
the kinetic and collisional transfer contributions to the pressure
tensor are evaluated. They are given as the computational analogs of Eqs.\ 
(\ref{p5377e4.0}) and (\ref{p5377e4}), i.e.
\begin{equation}
\label{p5377e14}
{\sf P}^k=\frac{mn}{N}\sum_{i=1}^{N} {\bf V}_i {\bf V}_i\; ,
\end{equation}
\begin{equation}
\label{p5377e15}
{\sf P}^c=\frac{mn}{N}\frac{\sigma}{\Delta t}{\sum_{ij}} ^{\dagger}
(\widehat{\mbox{\boldmath{$\sigma$}}}_{ij}\cdot {\bf g}_{ij})
\widehat{\mbox{\boldmath{$\sigma$}}}_{ij}
\widehat{\mbox{\boldmath{$\sigma$}}}_{ij}\; ,
\end{equation}
where the dagger means that the summation is restricted to the accepted
collisions.
Once the steady state is reached, the above quantities are averaged over 
time to improve the statistics.

\section{Results and Discussion}
Our  objective has been to obtain the pressure tensor ${\sf P}$ as a function
of the density and the shear rate 
by means of the kinetic model, as well as by performing 
Monte Carlo simulations of the EE. 

% Origin Apple Laser Writer Plus Vertical Page=8x8
\begin{figure}
\parbox{0.45\textwidth}{
\epsfxsize=\hsize \epsfbox{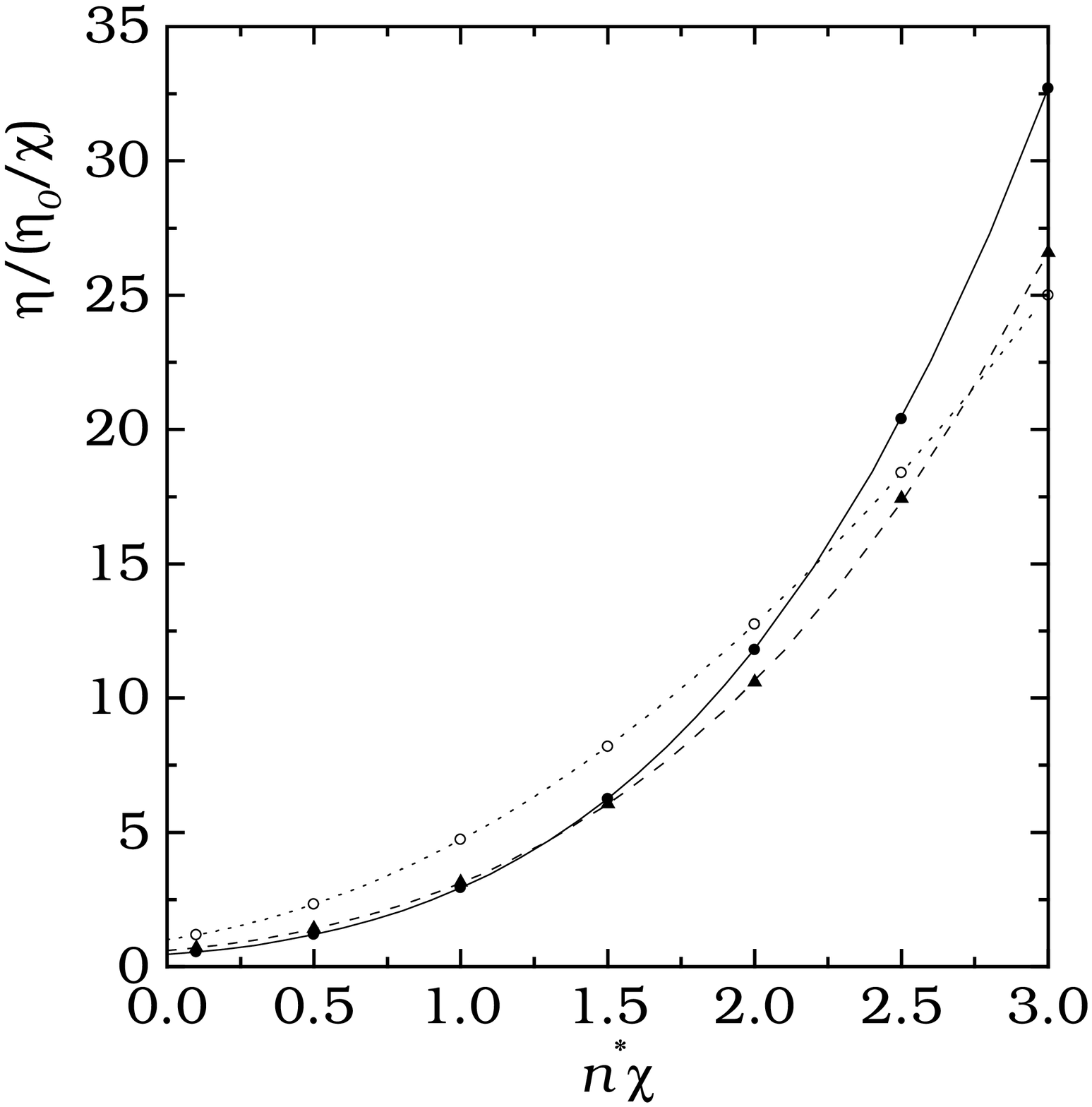}}
\hfill
\parbox{0.45\textwidth}{
\epsfxsize=\hsize \epsfbox{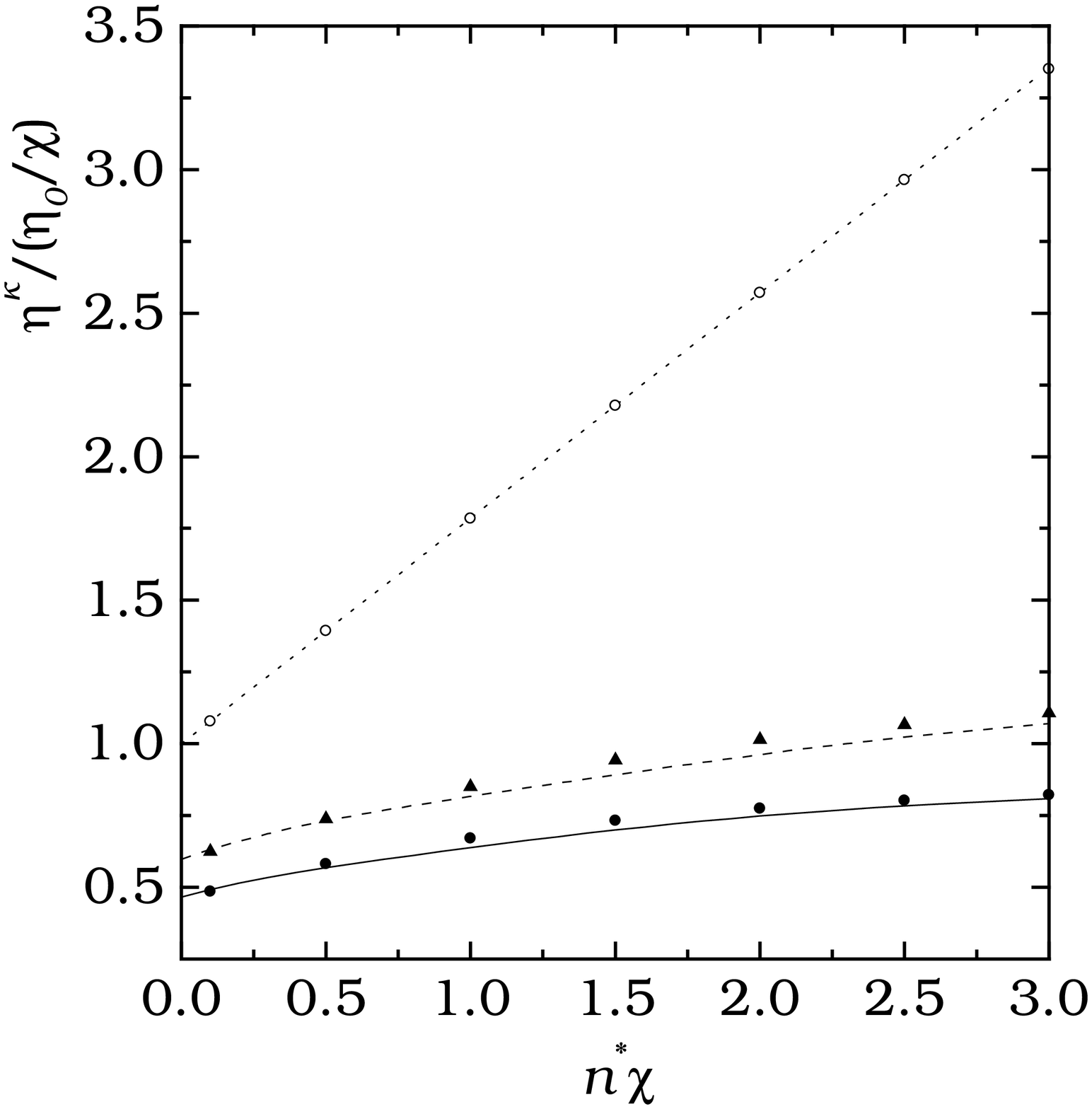}}
\caption{Plot of the normalized shear viscosity
$\eta/(\eta_0/\chi)$ and its kinetic part $\eta^k/(\eta_0/\chi)$
%as a function of the density parameter $n^*\chi$ 
for $a=0\ (\cdots)$, $a=0.7$ (-- --), 
and $a=1$ (---). Lines are from the kinetic model;
symbols are from Monte Carlo simulations of the EE.\label{p5377fig1}}
\end{figure}

Figure \ref{p5377fig1} shows a comparison of the (normalized)
non-Newtonian shear viscosity
$\eta=-P_{xy}/a$ as a function of the density parameter $n^*\chi$ for 
three different values of the shear rate.
The corresponding kinetic part is also shown in the right side 
of the figure. 
In all the figures presented in this paper, the
error bars are smaller than the sizes of the symbols and are not drawn.
The good agreement
indicates that both the kinetic and collisional transfer contributions
are accurately given by the model. 
The dotted lines correspond to the Navier-Stokes shear viscosity, Eq.\ 
(\ref{p5377e0}), so that the solid and the dashed lines represent 
non-Newtonian effects.
It is quite apparent that these effects are much more important at 
finite density
than at zero density. 
It is worthwhile  noticing that
the shear viscosity presents, in general, 
a non monotonic behavior as the shear rate increases.
More precisely, a transition from shear thinning, 
$\eta(n,a)<\eta_{\mbox{\scriptsize{NS}}}(n)$, 
to shear thickening,
$\eta(n,a)>\eta_{\mbox{\scriptsize{NS}}}(n)$,
takes place when the shear rate is larger than a critical value $a_c(n)$ 
that decreases as the density increases. In particular, $a_c=1.0$ at 
$n^*\chi\simeq 2.2$,
while $a_c=0.7$ at 
$n^*\chi\simeq 2.7$.
This transition
has been recently predicted for a hard-sphere fluid from an analysis of the 
kinetic model in the limit of small shear rates  and 
confirmed by the ESMC method \cite{p5377r6bis}.
This is an interesting effect not observed for the kinetic part.

\begin{figure}
\parbox{0.45\textwidth}{
\epsfxsize=\hsize \epsfbox{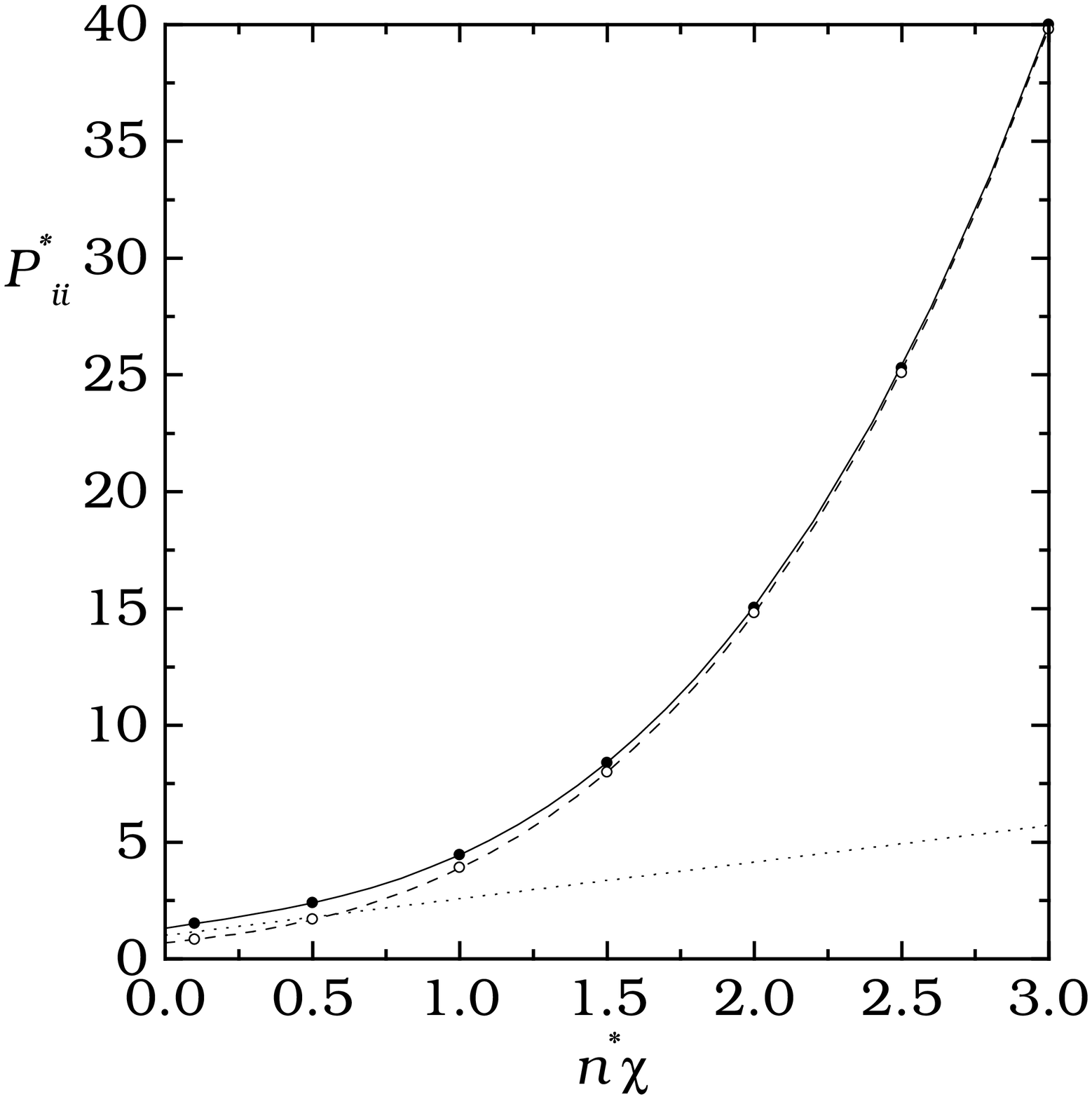}}
\hfill
\parbox{0.45\textwidth}{
\epsfxsize=\hsize \epsfbox{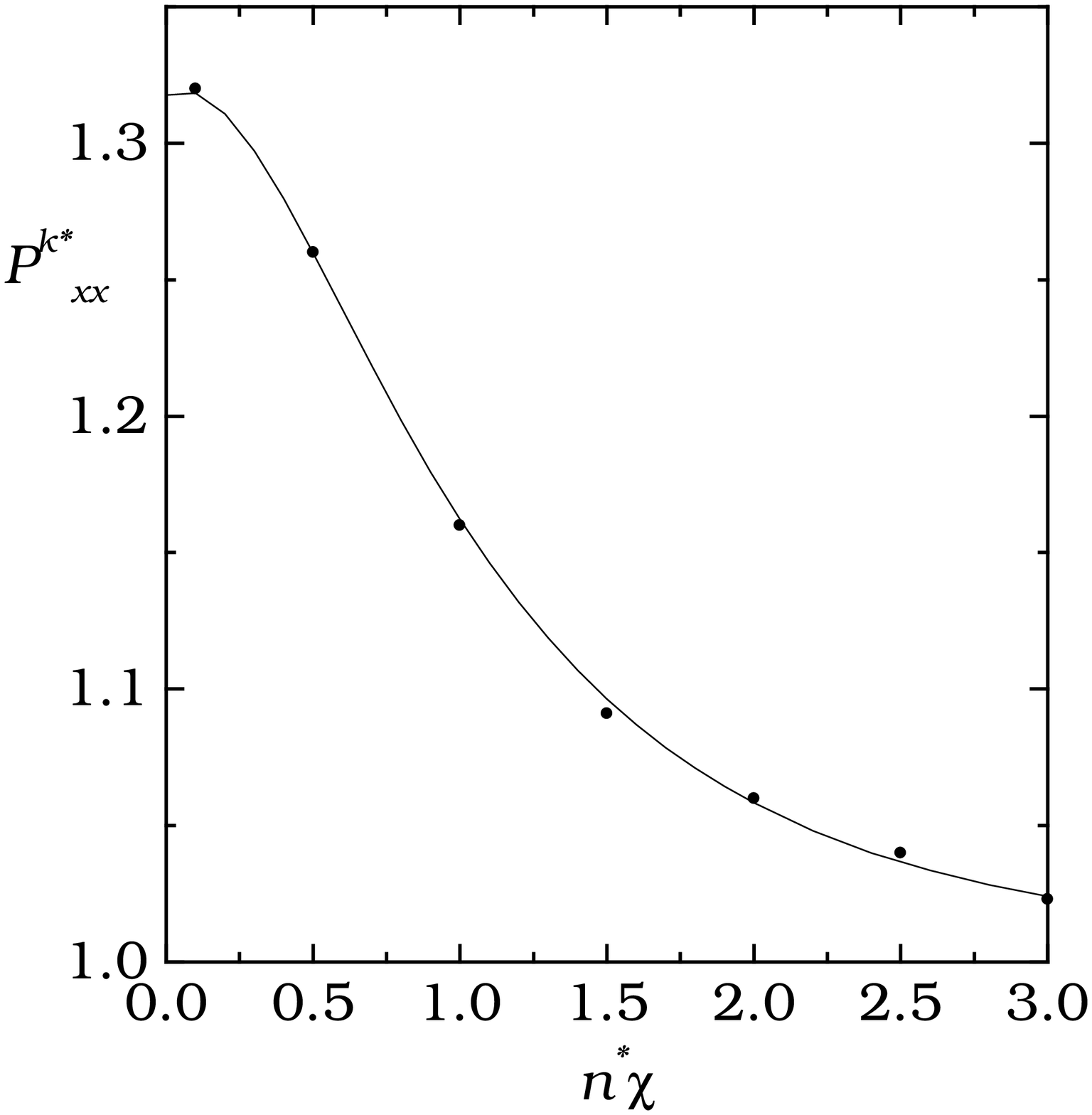}}
\caption{Plot of $P_{xx}^*$ (---), $P_{yy}^*$ (-- --) 
and $p_0^*$ ($\cdots$) (left side), 
and the kinetic part $P_{xx}^{k*}$ (right 
side), for $a=1$ . Lines are from the kinetic model; symbols are from Monte 
Carlo simulations of the EE.\label{p5377fig2}}
\end{figure}

Figure \ref{p5377fig2} shows the (dimensionless)
normal stresses $P_{xx}^*$ and $P_{yy}^*$ as 
functions of the density for the  shear rate  $a=1$. The dotted
line represents the equilibrium hydrostatic pressure 
$p_0^*=1+\frac{\pi}{2}  n^* \chi$ of a hard-disk fluid. 
Again, the numerical
solution of the model shows an excellent agreement for the wide
range of densities considered. 
At high densities the collisional part 
of the pressure tensor dominates and both diagonal elements $P^{*}_{xx}$ 
and $P^{*}_{yy}$ tend to coincide. 
In addition, viscometric effects become evident by observing
the increase of the nonequilibrium ``hydrostatic pressure"
$p^*=\frac{1}{2}\mbox{tr}{\sf P}^*$ with respect to its equilibrium
value $p_0^*$, a phenomenon usually referred to as ``shear dilatancy".
The kinetic contribution $P_{xx}^{k*}$ 
is also plotted in Fig.\ 2. 
Note that the $yy$-element can be
derived from the consistency condition $P_{xx}^{k*}+P_{yy}^{k*}=2$.
We observe that again a good accuracy of the model predictions 
holds for these quantities.

% Origin Apple Laser Writer Plus Vertical Page=8x8
\smallskip

In this work we have presented results obtained from both a kinetic
model \cite{p5377r6} 
and a simulation Monte Carlo method of the EE \cite{p5377r5} 
for a hard-disk system
under shear, arbitrarily far from equilibrium.
Although the EE is not expected to be accurate for very large densities,
we have  also considered those  densities in order to check the kinetic model
predictions. 
Comparison between the results obtained from the ESMC method  
and from the numerical solution of the model shows an excellent agreement, 
so that we can conclude that the kinetic model gives a good description of 
the
rheological properties throughout the whole shear rate-density plane.
In particular, we have analyzed  non-Newtonian effects 
beyond Navier-Stokes order by computing the elements of the pressure
tensor. A transition from shear thinning to shear 
thickening is clearly shown by using both approaches. 
\smallskip

Partial support from the DGICYT (Spain) through Grant No. PB97--1501  
and from the Junta de Extremadura (Fondo Social Europeo)
through Grant No. PRI97C1041 is acknowledged.


\begin{thebibliography}{99}
\bibitem{p5377r1}  Chapman S. and  Cowling T.G., 
\textit{The Mathematical Theory of Non-Uniform Gases},
Cambridge University Press, 1970.

\bibitem{p5377r2} Dufty J.W., 
\textit{Kinetic theory of fluids far from equilibrium --- Exact results}, in 
\textit{Lectures in Thermodynamics and Statistical Mechanics},
 L\'opez de Haro M. and  Varea C., eds., World Scientific, 
pp.166--181, 1990.

\bibitem{p5377r3}  Bird G.A., 
\textit{Molecular Gas Dynamics and the Direct Simulation of Gas Flows},
Clarendon Press, 1994.

\bibitem{p5377r4}  van Beijeren  H. and  Ernst M.H., \textit{The modified 
Enskog equation}, Physica,
Vol.68, pp.437--456, 1973.

\bibitem{p5377r5}  Montanero J.M. and  Santos A., 
\textit{Monte Carlo Simulation Method for the Enskog Equation},
Phys. Rev. E, Vol.54, No.1, pp.438--444, 1996; 
%Montanero J.M. and  Santos A.,
\textit{Viscometric Effects in a Dense Hard-Sphere Fluid}, Physica A,
Vol.240, Nos.1--2, pp.229--238, 1997.
\textit{Simulation of the Enskog Equation {\em \`a la} Bird}, Phys. Fluids, 
Vol.9, No.7, pp.2057--2060, 1997;
Frezzotti, A., \textit{A particle scheme for the numerical solution of the 
Enskog equation},
Phys. Fluids, Vol.9, No.5, pp.1329--1335.

\bibitem{p5377r6}  Dufty J.W.,  Santos A., and  Brey J.J.,
\textit{Practical Kinetic Model for Hard Sphere Dynamics},
Phys. Rev. Lett., Vol.77, No.7, pp.1270--1273, 1996; 
Dufty J.W.,   Brey J.J., and  Santos A.,
\textit{Kinetic models for hard sphere dynamics}, Physica A, Vol.240, 
Nos.1--2, pp.212--220, 1997.

\bibitem{p5377r6bis} 
Santos A.,  Montanero J.M.,
 Dufty J.W., and  Brey J.J., 
\textit{Kinetic Model for the hard-sphere fluid and solid},
Phys. Rev. E, Vol.57, No.2, pp.1644--1660, 1998.

%\bibitem{p5377r7} \textit{ Nonlinear Fluid Behavior}, 
%Hanley H.J.M., ed., North-Holland, 1983.

\bibitem{p5377r8}  Gass D.M., \textit{Enskog Theory for a Rigid Disk Fluid},
J. Chem. Phys., Vol.54, No.5, pp.1898--1902, 1971.

%\bibitem{p5377r9}  

\end{thebibliography}
\end{document}